\documentstyle[12pt]{article}

\def\be{\begin{equation}}
\def\ee{\end{equation}}
\def\bea{\begin{eqnarray}}
\def\eea{\end{eqnarray}}

\def\bar{\overline}

\def\a{\alpha}
\def\b{\beta}

\def\l{\lambda}
\def\bc{\begin{center}}
\def\ec{\end{center}}
\def\O{{\cal O}}
\def\sdpro{\mbox{
\begin{picture}(1,10)
\put(2,0){\line(0,1){6}}
\end{picture}
$\!\!\!\!\; \times$}}

\def\PR#1#2#3{Phys. Rev.  {\bf #1} (#3), #2}
\def\PRL#1#2#3{Phys. Rev. Lett. {\bf #1} (#3), #2}
\def\PL#1#2#3{Phys. Lett. B {\bf #1} (#3), #2}
\def\NP#1#2#3{Nucl. Phys. B {\bf #1} (#3), #2}
\def\PTP#1#2#3{Prog. Theor. Phys. {\bf #1} (#3), #2}
\begin{document} 

\begin{flushright}
hep-ph/0206232 \qquad AUE-02-01 / KGKU-02-01 / MIE-02-01 
\end{flushright}

\vspace{3mm}

\begin{center}
{\large \bf Non-Anomalous Flavor Symmetries \\
          and $SU(6) \times SU(2)_R$ Model }

\vspace{5mm}

Yoshikazu ABE$^1$, 
            \footnote{E-mail : abechan@phen.mie-u.ac.jp} 
Chuichiro HATTORI$^2$, 
            \footnote{E-mail : hattori@ge.aitech.ac.jp} 
Takemi HAYASHI$^3$, 
            \footnote{E-mail : hayashi@kogakkan-u.ac.jp} 
Masato ITO$^4$, 
            \footnote{E-mail : mito@eken.phys.nagoya-u.ac.jp} 
Masahisa MATSUDA$^5$, 
            \footnote{E-mail : mmatsuda@auecc.aichi-edu.ac.jp} 
Mamoru MATSUNAGA, 
            \footnote{E-mail : matsuna@phen.mie-u.ac.jp} 
and Takeo MATSUOKA$^6$, 
            \footnote{E-mail : matsuoka@kogakkan-u.ac.jp}

\end{center}

\begin{center}
{\it 
{}$^1$Department of Physics Engineering, Mie University, 
Tsu, 514-8507 JAPAN \\
{}$^2$Science Division, General Education, Aichi Institute of 
Technology, Toyota, 470-0392 JAPAN \\
{}$^3$Kogakkan University, Ise, 516-8555 JAPAN \\
{}$^4$Department of Physics, Nagoya University, Nagoya, 
464-8602 JAPAN \\
{}$^5$Department of Physics and Astronomy, Aichi University 
of Education, Kariya, 448-8542 JAPAN \\
{}$^6$Kogakkan University, Nabari, 518-0498 JAPAN 
}
\end{center}

\vspace{3mm}

\begin{abstract}
We introduce the flavor symmetry 
${\bf Z}_M \times {\bf Z}_N \times D_4$ 
into the $SU(6) \times SU(2)_{\rm R}$ string-inspired model. 
The cyclic group ${\bf Z}_M$ and the dihedral group $D_4$ are 
R symmetries, while ${\bf Z}_N$ is a non-R symmetry. 
By imposing the anomaly-free conditions on the model, 
we obtain a viable solution under many phenomenological 
constraints coming from the particle spectra. 
For the neutrino sector, we find a LMA-MSW solution but no 
SMA-MSW solution. 
The solution includes phenomenologically acceptable results 
concerning fermion masses and mixings and also concerning 
hierarchical energy scales including the GUT scale, 
the $\mu$ scale and the Majorana mass scale of R-handed neutrinos. 
\end{abstract}

\newpage 
\section{Introduction}
It is likely that in the framework of a unified theory, 
the characteristic patterns of fermion masses and mixings 
are closely linked to the flavor symmetry. 
In addition, it is feasible that the flavor symmetry also 
controls the GUT scale, the $\mu$ scale and the Majorana 
mass scale of R-handed neutrinos. 
In a previous paper\cite{Fuzz} the authors introduced 
the flavor symmetry ${\bf Z}_M \times {\bf Z}_N \times D_4$ 
into the $SU(6) \times SU(2)_{\rm R}$ string-inspired model, 
where ${\bf Z}_M$ and ${\bf Z}_N$ are R and ordinary symmetries, 
respectively. 
The dihedral group $D_4$ is also an R symmetry. 
The inclusion of $D_4$ is motivated by 
the phenomenological observation that the R-handed Majorana 
neutrino mass for the third generation is nearly equal to 
the geometrical average of the string scale $M_S$ 
and the electroweak scale $M_Z$. 
In the string theory it can be expected that the discrete flavor 
symmetries including the dihedral group $D_4$ arise from 
the symmetric structure of the compact space.

It has been pointed out that all non-gauge symmetries are strongly 
violated by quantum gravity effects around the Planck scale 
and hence in the low-energy effective theory we cannot have 
any global symmetries.\cite{Banks} 
This statement holds even for the discrete symmetry 
introduced above. 
In contrast to the situation for non-gauge symmetries, 
if the flavor symmetries are unbroken discrete subgroups of 
local gauge symmteries, 
the discrete flavor symmetries are stable with respect to 
quantum gravity effects and therefore remain in the low-energy 
effective theory. 
Such discrete flavor symmetries are subject to certain 
anomaly cancellation conditions.\cite{Ib-Ro,KMY} 
These conditions are so stringent that many candidates of 
discrete symmetries are ruled out. 
Although in Ref.\cite{Fuzz} the authors found interesting 
solutions that yield not only fermion mass hierarchies 
but also hierarchical energy scales, 
the flavor symmetry adopted there is 
inconsistent with the anomaly-free conditions. 
The purpose of this paper is to explore the non-anomalous 
flavor symmetry ${\bf Z}_M \times {\bf Z}_N \times D_4$ 
and to find phenomenologically viable anomaly free solutions.

This paper is organized as follows. 
In section 2 we briefly explain the main features of 
the $SU(6) \times SU(2)_{\rm R}$ string-inspired model, 
in which ${\bf Z}_M \times {\bf Z}_N$ and the dihedral group 
$D_4$ symmetries are introduced as the flavor symmetry. 
We use a projective representation of $D_4$, which is expected to 
arise in the theory on a compact space with non-commutative 
geometry. 
It is pointed out that the $D_4$ symmetry is an extension of 
the R-parity. 
In section 3 we study phenomenological constraints on the flavor 
charges of the matter fields. 
These constraints come from fermion mass hierachies 
and mixings and also from hierarchical energy scales. 
The anomaly-free conditions are given in section 4. 
Important conditions arise from the flavor-gauge mixed anomalies. 
In section 5 we solve the anomaly-free conditions, taking account of 
the phenomenological constraints and present a large mixing 
angle (LMA)-MSW solution. 
However, small mixing angle (SMA)-MSW solutions could not be 
found in the region of plausible parameter values. 
The distinction between these solutions results from the 
difference in the flavor charge assignments. 
We obtain phenomenologically viable results regarding fermion 
masses and mixings and also regarding hierarchical energy scales, 
including the GUT scale, the $\mu$ scale and the Majorana 
mass scale of R-handed neutrinos. 
The final section is devoted to summary and discussion.

\vspace{10mm}

\section{$SU(6) \times SU(2)_{\rm R}$ Model}
The $SU(6) \times SU(2)_{\rm R}$ string-inspired model considered here 
is studied in detail in 
Refs. \cite{Matsu1,Matsu2,Matsu3,CKM,MNS}. 
In this section we review the main features of the model. 
\begin{enumerate}
\item The unified gauge symmetry $G$ at the string scale $M_S$ 
is assumed to be $SU(6) \times SU(2)_{\rm R}$. 

\item Matter consists of chiral superfields of three families and one 
vector-like multiplet, i.e., 
\begin{equation}
  3 \times {\bf 27}(\Phi_{1,2,3}) + 
        ({\bf 27}(\Phi_0)+\overline{\bf 27}({\bar \Phi})), 
\end{equation}
in terms of $E_6$. 
The superfields $\Phi$ in {\bf 27} of $E_6$ are decomposed into 
irreducible representations of $G = SU(6) \times SU(2)_{\rm R}$ as 
\begin{equation}
  \Phi({\bf 27})=\left\{
       \begin{array}{lll}
         \phi({\bf 15},{\bf 1})& : 
               & \quad \mbox{$Q,L,g,g^c,S$}, \\
          \psi({\bf 6}^*,{\bf 2}) & : 
               & \quad \mbox{$(U^c,D^c),(N^c,E^c),(H_u,H_d)$}, 
       \end{array}
       \right.
\label{eqn:27}
\end{equation}
where the pair $g$ and $g^c$ and the pair $H_u$ and $H_d$ represent 
colored Higgs and doublet Higgs superfields, respectively, 
$N^c$ is the right-handed neutrino superfield, and 
$S$ is an $SO(10)$ singlet. 

\item Gauge invariant trilinear couplings in the superpotential 
$W$ take the forms 
\bea
    (\phi ({\bf 15},{\bf 1}))^3 & = & QQg + Qg^cL + g^cgS, \\
    \phi ({\bf 15},{\bf 1})(\psi ({\bf 6}^*,{\bf 2}))^2 & 
            = & QH_dD^c + QH_uU^c + LH_dE^c  + LH_uN^c 
                                            \nonumber \\ 
             {}& & \qquad   + SH_uH_d + 
                     gN^cD^c + gE^cU^c + g^cU^cD^c. 
\eea 
\end{enumerate}

The gauge group $G = SU(6) \times SU(2)_{\rm R}$ can be obtained from 
$E_6$ through the Hosotani mechanism or flux breaking 
on multiply-connected manifolds.\cite{Hoso,Flux1,Flux2} 
We construct the multiply-connected manifold $K$ as the 
coset $K_0/G_d$ of a simply-connected $K_0$ modded out 
by a discrete group $G_d$ of $K_0$. 
In the presence of a background gauge field for 
extra-dimensional components, we have a nontrivial holonomy 
$U_d$ on $K = K_0/G_d$. 
This nontrivial $U_d$ gives rise to the discrete symmetry 
${\overline G_d}$, which is an embedding of $G_d$ into $E_6$. 
The unbroken gauge group $G$ is the subgroup of $E_6$ whose 
elements commute with all elements of ${\overline G_d}$. 
When the holonomy $U_d$ is of the form 
\begin{equation}
   U_d = \exp( \pi i I_3(SU(2))), 
\end{equation}
we obtain ${\overline G_d} \equiv {\bf Z}_2^{({\rm W})}$, 
where $I_3$ represents the third direction of 
an appropriate $SU(2)$ in $E_6$. 
The gauge group $G$ becomes $SU(6) \times SU(2)$.\cite{Matsu4} 
The superfield {\bf 27} of $E_6$ is decomposed into two 
irreducible representations $\phi({\bf 15, \ 1})$ and 
$\psi({\bf 6^*, \ 2})$, 
which are even and odd under ${\bf Z}_2^{({\rm W})}$ 
parity, respectively.

In the conventional GUT-type models, 
unless an adjoint or higher representation matter (Higgs) field 
develops a non-zero VEV, 
it is impossible for a large gauge symmetry such as $SU(5)$ 
or $SO(10)$ to be spontaneously broken down to the standard 
model gauge group $G_{\rm SM}$ via the Higgs mechanism. 
Contrastingly, in the present model, matter fields consist only 
of ${\bf 27}$ and $\overline{\bf 27}$. 
The symmetry breaking of $G = SU(6) \times SU(2)_{\rm R}$ 
down to $G_{\rm SM}$ can take place via the Higgs mechanism without 
matter fields of adjoint or higher representations. 
In addition, $SU(6) \times SU(2)_{\rm R}$ is the largest of 
such gauge groups. 
Furthermore, it should be noted that doublet Higgs and color-triplet 
Higgs fields belong to different irreducible representations 
of $G$, as shown in Eq. (\ref{eqn:27}). 
As a consequence, the triplet-doublet splitting problem is 
solved naturally.\cite{Matsu1}

As the flavor symmetry, we introduce 
${\bf Z}_M \times {\bf Z}_N$ and $D_4$ symmetries 
and regard ${\bf Z}_M$ and ${\bf Z}_N$ as the R and 
non-R symmetries, respectively. 
Assuming that $M$ and $N$ are relatively prime, 
we combine these symmetries as 
\begin{equation}
  {\bf Z}_M \times {\bf Z}_N = {\bf Z}_{MN}. 
\end{equation}
In this case we stipulate that the Grassmann number $\theta$ 
in the superfield formalism has the charge $(\pm 1, \ 0)$ 
under ${\bf Z}_M \times {\bf Z}_N$. 
The charge of $\theta$ under ${\bf Z}_{MN}$ is denoted as 
$q_{\theta}$, which becomes a multiple of $N$, and 
$q_{\theta} \equiv \pm 1$ (mod $M$). 
The ${\bf Z}_{MN}$ charges of matter superfields are denoted 
as $a_i$ and $b_i$, etc., as shown in Table I.

\begin{table}[t]
\caption{Assignment of ${\bf Z}_{MN}$ charges 
          for matter superfields}
\label{table:I}
\bc
\begin{tabular}{|c|ccc|} \hline \hline 
\vphantom{\LARGE I} & \phantom{M} $\Phi_i \ (i=1,2,3)\ $ \phantom{M} & 
        \phantom{M} $\Phi_0$ \phantom{M} & 
              \phantom{M} $\bar{\Phi}$ \phantom{M} \\ \hline
$\phi({\bf 15, \ 1})$   &  $a_i$  &  $a_0$  &  $\bar{a}$  \\
$\psi({\bf 6^*, \ 2})$  &  $b_i$  &  $b_0$  &  $\bar{b}$  \\ \hline
\end{tabular}
\ec
\end{table}

Introduction of the dihedral group 
$D_4={\bf Z}_2 \sdpro {\bf Z}_4$ 
is motivated by the phenomenological observation that 
the R-handed Majorana neutrino mass for the third generation 
is nearly equal to the geometrical average of $M_S$ and $M_Z$. 
The ${\bf Z}_2$ and ${\bf Z}_4$ groups are expressed as 
\begin{equation}
  {\bf Z}_2 = \{1, \ g_1 \}, \qquad 
  {\bf Z}_4 = \{1, \ g_2, \ g_2^2, \ g_2^3 \}, 
\end{equation}
respectively, and we have the relation 
\begin{equation}
   g_1 g_2 g_1^{-1} = g_2^{-1}. 
\end{equation}
The elements  $g_1$ and $g_2$ correspond to reflection and 
rotation by $\pi/2$ of a square, respectively.

The reader might think that the $D_4$ symmetry is somewhat 
unfamiliar as the flavor symmetry. 
However, examples of $D_4$ symmetric Calabi-Yau space can be 
easily constructed as follows. 
We first note that zero locus of the 5th-order defining polynomial 
in $CP^4$ is a simple example of the Calabi-Yau space. 
Denoting the homogeneous coordinates of $CP^4$ as $z_i$ 
$(i=1, \ 2, \ \cdots 5)$, 
we take the defining polynomial as 
\begin{equation}
   P(z) = \sum_{i=1}^5 z_i^5 + c z_5^3 (z_1 z_3 + z_2 z_4), 
\end{equation}
where $c$ is a complex constant. 
The defining polynomial $P(z)$ is invariant under the 
transformation 
\begin{equation}
    g_1 \ : \ z_1 \leftrightarrow z_3, \quad 
      z_i \rightarrow z_i, \ \ (i=2,4,5) 
\end{equation}
which composes ${\bf Z}_2^{({\rm A})} = \{1, \ g_1 \}$, 
and also under the transformation 
\begin{equation}
    g_2 \ : \ z_1 \rightarrow z_2 \rightarrow z_3 \rightarrow 
     z_4 \rightarrow z_1, \quad z_5 \rightarrow z_5, 
\end{equation}
which composes ${\bf Z}_4 = \{1, \ g_2, \ g_2^2, \ g_2^3 \}$. 
These transformations yield the dihedral group 
$D_4={\bf Z}_2^{({\rm A})} \sdpro {\bf Z}_4$. 
Then, $D_4$ symmetry arises on the compact space constructed here. 
This simple example suggests that it is not so unusual that 
the dihedral group $D_4$ is included among the flavor symmetries 
in the effective theory from the string compactification.

Furthermore, when $c$ is real, instead of the above 
${\bf Z}_2^{({\rm A})}$ transformation, we may adopt another 
${\bf Z}_2$ transformation, 
\begin{equation}
   g'_1 \ : \ z_1 \rightarrow \bar{z_3}, \quad 
                    z_3 \rightarrow \bar{z_1}, \quad 
                    z_i \rightarrow \bar{z_i}, \ \ (i=2,4,5) 
\end{equation}
which is a combined transformation ${\bf Z}_2^{({\rm AC})}$ 
consisting of ${\bf Z}_2^{({\rm A})}$ and complex conjugation. 
The operation of complex conjugation corresponds to the reversal 
of the string orientation. 
Under this transformation, $P(z)$ transforms into 
$P(\bar{z}) = \bar{P(z)}$. 
Then, the defining polynomial remains essentially unchanged. 
Although chiral matter superfields transform into anti-chiral ones, 
the terms coming from the superpotential 
\begin{equation}
   \int d \theta^2 \, W + \int d \bar{\theta}^2 \, \bar{W} 
\end{equation}
are invariant under the ${\bf Z}_2^{({\rm AC})}$ transformation, 
provided that $\theta \ (\bar{\theta})$ transforms into 
$\bar{\theta} \ (\theta)$ simultaneously.

It is assumed that the flavor symmetry contains 
the dihedral group $D_4$. 
Here we denote this $D_4$ as 
${\bf Z}_2^{({\rm F})} \sdpro {\bf Z}_4$. 
In a string with discrete torsion, the coordinates in the 
compact space become non-commutative and are 
represented by a projective representation of the flavor 
symmetry.\cite{torsion,torsion2,B-f} 
This non-commutativity of the coordinates corresponds to 
brane fluctuations. 
The non-commutative coordinates are concretely represented 
in terms of matrices.\cite{Beren} 
Massless matter fields in the effective theory correspond to 
the degree of freedom of deformation of the compact space 
and are expressed by functions of non-commutative coordinates. 
Therefore, massless matter fields turn out to be of matrix form. 
Specifically, the matter fields are described in terms of 
the ordinary four-dimensional fields mutiplied by the matrices 
associated with the non-commutativity of the compact space. 
The four-dimensional Lagrangian of the theory should belong 
to the center of the non-commutative algebra. 
As pointed out in Ref.\cite{Fuzz}, 
this implies that a new type of flavor symmetry arises in the 
theory on a compact space with non-commutative geometry.

To begin with, let us consider a projective representation of 
the dihedral group 
$D_4={\bf Z}_2^{({\rm F})} \sdpro {\bf Z}_4$. 
It is easy to see that a projective representation of this $D_4$ 
is given by the unitary matrices 
\begin{equation}
  \gamma(g_1) = \left(
       \begin{array}{cc}
         0  &  1  \\
         1  &  0  
       \end{array}
       \right) = \sigma_1, \qquad 
  \gamma(g_2) = \left(
       \begin{array}{cc}
         1  &  0  \\
         0  &  i  
       \end{array}
       \right) \equiv \sigma_4, 
\label{eqn:prjrep}
\end{equation}
which satisfy the relations 
\begin{equation}
   \gamma(g_1) \, \gamma(g_2) \, \gamma(g_1)^{-1} 
                            = i \, \gamma(g_2)^{-1}, 
\qquad \gamma(g_1)^2 = \gamma(g_2)^4 = 1. 
\end{equation}
In this case we have 
\begin{equation}
  \gamma(g_1 \, g_2^2) = \left(
       \begin{array}{cc}
         0  &  -1  \\
         1  &   0  
       \end{array}
       \right) = -i \sigma_2, \qquad 
  \gamma(g_2^2) = \left(
       \begin{array}{cc}
         1  &   0  \\
         0  &  -1  
       \end{array}
       \right) = \sigma_3. 
\end{equation}
In $D_4$ there exist five conjugacy classes, 
\begin{equation}
  \{ 1 \}, \quad \{ g_1, \ g_1 g_2^2 \}, \quad \{ g_2^2 \}, \quad 
       \{ g_2, \ g_2^3 \}, \quad \{ g_1 g_2, \ g_1 g_2^3 \}. 
\end{equation}
Correspondingly, for example, $1$ and $\sigma_i \ (i=1,2,3,4)$ 
transform as 
\bea
   \gamma(g_1) \, \{ 1, \ \sigma_1, \ \sigma_2, \ 
       \sigma_3, \ \sigma_4 \} \, \gamma(g_1)^{-1} & = & 
                 \{ 1, \ \sigma_1, \ -\sigma_2, \ 
                               -\sigma_3, \ i\sigma_4^{-1} \}, \\
   \gamma(g_2) \, \{ 1, \ \sigma_1, \ \sigma_2, \ 
       \sigma_3, \ \sigma_4 \} \, \gamma(g_2)^{-1} & = & 
                 \{ 1, \ \sigma_2, \ -\sigma_1, \ 
                               \sigma_3, \ \sigma_4 \}. 
\eea
To each matter superfield we assign a ``$D_4$-charge" 
which is expressed in terms of the representation matrices 
of $D_4$.

We now define a combined transformation ${\bf Z}_2^{({\rm FC})}$ 
consisting of ${\bf Z}_2^{({\rm F})}$ and hermitian conjugation. 
In addition, we define a combined transformation consisting of 
${\bf Z}_2^{({\rm FC})}$ and ${\bf Z}_2^{({\rm W})}$ by 
${\bf Z}_2^{({\rm FCW})}$ and require that the theory be 
${\bf Z}_2^{({\rm FCW})}$ gauge-invariant. 
This means that the theory on a manifold $K_0$ is modded out by 
the combined ${\bf Z}_2^{({\rm FCW})}$. 
Because the field $\phi({\bf 15, \ 1})$ is even under 
${\bf Z}_2^{({\rm W})}$, ${\bf Z}_2^{({\rm FC})}$ odd states 
are projected out for $\phi({\bf 15, \ 1})$. 
Then the representation of $D_4$ for the field $\phi({\bf 15, \ 1})$ 
is 1 or $\sigma_1$. 
This situation is described in Table II. 
In this table, $\sigma_4$ is redefined by attaching 
the phase factor $\exp(i\pi/4)$. 
Then, $\sigma_4$ and $\sigma_4^{-1}$ become odd under 
${\bf Z}_2^{({\rm FC})}$. 
On the other hand, since $\psi({\bf 6^*, \ 2})$ is odd under 
${\bf Z}_2^{({\rm W})}$, 
${\bf Z}_2^{({\rm FC})}$ even states are projected out 
for $\psi({\bf 6^*, \ 2})$. 
Then the representation of $D_4$, i.e. $\sigma_2$, $\sigma_3$, 
$\sigma_4$ or $\sigma_4^{-1}$, is attached to the field 
$\psi({\bf 6^*, \ 2})$. 
The disappearance of $\sigma_1$ ($\sigma_2$) from the spectra 
of $\psi({\bf 6^*, \ 2})$ ($\phi({\bf 15, \ 1})$) induces 
the breakdown of 
$D_4 = {\bf Z}_2^{({\rm F})} \sdpro {\bf Z}_4$ 
to ${\bf Z}_2^{({\rm F})} \times {\bf Z}_2$.

\begin{table}
\caption{${\bf Z}_2$ parities of matter superfields}
\label{table:II}
\bc
\begin{tabular}{|c|ccc|} \hline \hline 
 \vphantom{\LARGE I} & ${\bf Z}_2^{({\rm W})}$  
                         &  ${\bf Z}_2^{({\rm FC})}$  
                             &  ${\bf Z}_2^{({\rm FCW})}$    \\ \hline
$({\bf 15, \ 1}) \ \,  1 \ $       &   $+$   &   $+$  &   $+$   \\
$({\bf 15, \ 1}) \ \sigma_1$       &   $+$   &   $+$  &   $+$   \\
$({\bf 15, \ 1}) \ \sigma_2$       &   $+$   &   $-$  &   $-$   \\
$({\bf 15, \ 1}) \ \sigma_3$       &   $+$   &   $-$  &   $-$   \\
$({\bf 15, \ 1}) \ \sigma_4$       &   $+$   &   $-$  &   $-$   \\ 
$({\bf 15, \ 1}) \ \sigma_4^{-1}$  &   $+$   &   $-$  &   $-$   \\ \hline
$({\bf 6^*, \ 2}) \ \,  1 \ $      &   $-$   &   $+$  &   $-$   \\
$({\bf 6^*, \ 2}) \ \sigma_1$      &   $-$   &   $+$  &   $-$   \\
$({\bf 6^*, \ 2}) \ \sigma_2$      &   $-$   &   $-$  &   $+$   \\
$({\bf 6^*, \ 2}) \ \sigma_3$      &   $-$   &   $-$  &   $+$   \\
$({\bf 6^*, \ 2}) \ \sigma_4$      &   $-$   &   $-$  &   $+$   \\
$({\bf 6^*, \ 2}) \ \sigma_4^{-1}$ &   $-$   &   $-$  &   $+$   \\  \hline
\end{tabular}
\ec
\end{table}

\begin{table}
\caption{Assignment of ``$D_4$ charges" to matter superfields}
\label{table:III}
\bc
\begin{tabular}{|c|ccc|} \hline \hline 
\vphantom{\LARGE I} & \phantom{M} $\Phi_i \ (i=1,2,3)\ $ \phantom{M} & 
        \phantom{M} $\Phi_0$ \phantom{M} & 
              \phantom{M} $\bar{\Phi}$ \phantom{M} \\ \hline
$\phi({\bf 15, \ 1})$   & 
                   $\sigma_1$  &     1       &       1       \\
$\psi({\bf 6^*, \ 2})$  & 
                   $\sigma_2$  & $\sigma_3$  &   $\sigma_4$  \\  \hline
\end{tabular}
\ec
\end{table}

\begin{table}
\caption{R-parities of matter superfields}
\label{table:IV}
\bc
\begin{tabular}{|c|ccc|} \hline \hline 
\vphantom{\LARGE I} & \phantom{M} $\Phi_i \ (i=1,2,3)\ $ \phantom{M} & 
        \phantom{M} $\Phi_0$ \phantom{M} & 
              \phantom{M} $\bar{\Phi}$ \phantom{M} \\ \hline
$\phi({\bf 15, \ 1})$   &  $-$   &   $+$   &   $+$   \\
$\psi({\bf 6^*, \ 2})$  &  $-$   &   $+$   &   $+$   \\  \hline
\end{tabular}
\ec
\end{table}

We are now in a position to assign the ``$D_4$-charges" to matter fields, 
as shown in Table III, and $\sigma_1$ to the Grassmann number $\theta$. 
It is worth noting that the $\sigma_3$ transformation yields 
the R-parity. 
In fact, we find that 
\begin{equation}
  \sigma_3 \sigma_1 \sigma_3^{-1} = - \sigma_1, \qquad 
  \sigma_3 \sigma_2 \sigma_3^{-1} = - \sigma_2
\end{equation}
and 
\begin{equation}
  \sigma_3 1 \sigma_3^{-1} = 1, \qquad 
  \sigma_3 \sigma_4 \sigma_3^{-1} = \sigma_4, \qquad 
  \sigma_3 \sigma_3 \sigma_3^{-1} = \sigma_3. 
\end{equation}
In other words, the R-parities of the superfields $\Phi_i \, (i=1,2,3)$ 
for three generations are all odd, 
while those of $\Phi_0$ and $\bar{\Phi}$ are even. 
This is shown in Table IV. 
Therefore, the dihedral flavor symmetry $D_4$ is 
an extension of the R-parity. 
When $\psi({\bf 6^*, \ 2})_0$ and $\bar{\psi({\bf 6^*, \ 2})}$ 
develop non-zero VEVs, 
the ${\bf Z}_2^{({\rm F})}$ symmetry is spontaneously broken. 
Eventually, the dihedral flavor symmetry 
$D_4={\bf Z}_2^{({\rm F})} \sdpro {\bf Z}_4$ is 
spontaneously broken down to ${\bf Z}_2^{({\rm R})}$ symmetry. 
This ${\bf Z}_2^{({\rm R})}$ symmetry is a remnant of 
the ${\bf Z}_4$ symmetry and is identified with the R-parity.

\vspace{10mm}

\section{Fermion mass hierarchies and mixings}
In this section we study phenomenological requirements, 
which yield many constraints on the assignments of the discrete 
flavor charges. 
Our purpose is to explain not only the fermion mass hierarchies 
and the mixings but also the hierachical energy scales, 
including the breaking scale of the GUT-type gauge symmetry, 
the intermediate Majorana masses of the R-handed neutrinos and 
the scale of the $\mu$ term.

In the R-parity even sector, it is assumed that 
the superpotential contains the terms 
\begin{equation}
  W_1 \sim M_S^3 \left[ \l_0 
     \left( \frac{\phi_0 \bar{\phi}}{M_S^2} \right)^{2n} 
       + \l_1 \left( \frac{\phi_0 \bar{\phi}}{M_S^2} \right)^n 
         \left( \frac{\psi_0 \bar{\psi}}{M_S^2} \right)^m 
       + \l_2 \left( \frac{\psi_0 \bar{\psi}}{M_S^2} \right)^{2m}\right], 
\label{eqn:W1}
\end{equation}
with $\l_i = \O(1)$, where the exponents are non-negative integers that 
satisfy the ${\bf Z}_{MN}$ symmetry conditions, 
\bea
  2n (a_0 + \bar{a}) - 2 q_{\theta} & \equiv & 0,  \nonumber \\
  n (a_0 + \bar{a}) + m (b_0 + \bar{b}) 
           - 2 q_{\theta} & \equiv & 0, \qquad ({\rm mod} \ MN) \\
  2m (b_0 + \bar{b}) - 2 q_{\theta} & \equiv & 0. \nonumber
\label{eqn:fcm1}
\eea
The dihedral symmetry $D_4$ requires $m \equiv 0$ (mod 4). 
Then, for the sake of simplicity we put $m=4$. 
As discussed in Ref.\cite{Fuzz}, we consider the case that $M$ is odd and 
$N \equiv 2$ (mod 4). 
Furthermore, the ${\bf Z}_{MN}$ charges are chosen as 
\begin{equation}
  a_0 + \bar{a} = -4, \qquad \bar{b} = {\rm odd}, \qquad 
     a_i, \ b_i = {\rm even}, \ (i=0,1,2,3) 
\label{eqn:ab1}
\end{equation}
and 
\begin{equation}
  a_i + a_j, \ a_0, \ \bar{a}, \  b_i + b_j \equiv 0. 
                                \qquad ({\rm mod} \ 4) 
\label{eqn:ab2}
\end{equation}
In this case we obtain 
\begin{equation}
  n = \frac{1}{4} (MN - q_{\theta}) = - (b_0 + \bar{b}). 
\label{eqn:fc1}
\end{equation}
Through the minimization of the scalar potential with 
the soft SUSY breaking mass terms characterized by 
the scale $\widetilde{m}_0 \sim 10^3$ GeV, 
matter fields develop non-zero VEVs. 
In Refs. \cite{Scale1} and \cite{Scale2} we studied 
the minimum point of the scalar potential in detail. 
The gauge symmetry is spontaneously broken in two steps 
with a feasible parameter region of the coefficients 
$\l_i$. 
The scales of the gauge symmetry breaking are given by 
\bea
  |\langle \phi_0 \rangle| = |\langle \bar{\phi} \rangle| 
        & = & M_S \, \rho^{1/2(2n-1)},  \nonumber  \\
  |\langle \psi_0 \rangle| = |\langle \bar{\psi} \rangle| 
        & \simeq & M_S \, \rho^{n/8(2n-1)}. 
\label{eqn:scale}
\eea
The parameter $\rho$ is defined by 
$\rho = c \, \widetilde{m}_0/M_S$, 
where $c \simeq n^{-3/2}\, f(\l_0, \, \l_1, \, \l_2)$. 
Here we take the numerical values 
$M_S \sim 5 \times 10^{17}$GeV\cite{MS} 
and $c \sim 10^{-2}$. 
Thus, we have $\rho \sim 2 \times 10^{-17}$. 
The $D$-flat conditions require 
$|\langle \phi_0 \rangle| = |\langle \bar{\phi} \rangle|$ and 
$|\langle \psi_0 \rangle| = |\langle \bar{\psi} \rangle|$ 
with $\O(M_S \, \rho)$ accuracy. 
Under the assumption $n > m = 4$, we have 
\begin{equation}
   |\langle \phi_0 \rangle| > |\langle \psi_0 \rangle|. 
\end{equation}
In what follows, we use the notation 
\begin{equation}
   \frac {\langle \phi_0 \rangle \langle {\overline \phi} \rangle}{M_S^2} = x, 
   \qquad 
   \frac {\langle \psi_0 \rangle \langle {\overline \psi} \rangle}{M_S^2} 
           = x^{\frac{n}{4} + \delta_N}, 
\end{equation}
with $x^{\delta_N} \sim 1$. 
Then we have 
\begin{equation}
  x^{2n-1} = \rho \sim 2 \times 10^{-17}. 
\end{equation}

The gauge symmetry is spontaneously broken 
at the scale $|\langle \phi_0({\bf 15, \ 1})\rangle |$, and 
subsequently at the scale 
$|\langle \psi_0({\bf 6^*, \ 2})\rangle |$. 
This yields the symmetry breakings 
\begin{equation}
   SU(6) \times SU(2)_{\rm R} 
     \buildrel \langle \phi_0 \rangle \over \longrightarrow 
             SU(4)_{\rm PS} \times SU(2)_L \times SU(2)_{\rm R}  
     \buildrel \langle \psi_0 \rangle \over \longrightarrow 
     G_{\rm SM}, 
\end{equation}
where $SU(4)_{\rm PS}$ is the Pati-Salam $SU(4)$.\cite{Pati} 
Since the fields that develop non-zero VEVs are singlets under 
the remaining gauge symmetries, 
they are assigned as 
$\langle \phi_0({\bf 15, \ 1})\rangle = \langle S_0 \rangle $ and 
$\langle \psi_0({\bf 6^*, \ 2})\rangle = \langle N^c_0 \rangle $.
Below the scale $|\langle \phi_0 \rangle|$, 
the Froggatt-Nielsen mechanism acts for non-renormalizable 
interactions.\cite{F-N} 
In the first step of the symmetry breaking, the fields $Q_0$, $L_0$, 
${\overline Q}$, ${\overline L}$ and $(S_0 - {\overline S})/\sqrt{2}$ 
are absorbed by the gauge fields. 
Through subsequent symmetry breaking, the fields $U_0^c$, $E_0^c$, 
${\overline U}^c$, ${\overline E}^c$ and 
$(N_0^c - {\overline N}^c)/\sqrt{2}$ are absorbed.

The colored Higgs mass arises from the term 
\begin{equation}
  z_{00} \left( \frac {S_0 {\overline S}}{M_S^2} \right)^{\zeta _{00}} 
     S_0 g_0 g^c_0, 
\label{eqn:cHiggs}
\end{equation}
with $z_{00} = \O(1)$. 
The ${\bf Z}_{MN}$ symmetry controls the exponent $\zeta_{00}$ as 
\begin{equation}
  -4 \zeta_{00} + 3a_0 - 2 q_{\theta} \equiv 0, 
                                            \qquad ({\rm mod} \ MN) 
\label{eqn:fcm2}
\end{equation}
where we have used $a_0 + \bar{a} = -4$. 
Due to the Froggatt-Nielsen mechanism, 
the colored Higgs mass can be expressed as 
\begin{equation}
  m_{g_0/g^c_0} \sim x^{\zeta_{00}} \, \langle S_0 \rangle. 
\end{equation}
In order to guarantee the longevity of the proton, 
$\zeta_{00}$ should be sufficiently small compared to $n$. 
For this reason, we rewrite Eq.~(\ref{eqn:fcm2}) as 
\begin{equation}
  -4 \zeta_{00} + 3a_0 - 2 q_{\theta} = 0, 
\label{eqn:fc2}
\end{equation}
which gives a small non-negative value of $\zeta_{00}$ when 
$3a_0 - 2q_{\theta}$ is a small non-negative multiple of 4.

Similarly, the $\mu$ term induced from 
\begin{equation}
   h_{00} \left( \frac {S_0 {\overline S}}{M_S^2} \right)^{\eta _{00}} 
       S_0 H_{u0} H_{d0}, 
\label{eqn:mu1}
\end{equation}
with $h_{00} = \O(1)$, is of the form 
\begin{equation}
  \mu \sim x^{\eta_{00}} \, \langle S_0 \rangle. 
\end{equation}
The exponent $\eta_{00}$ is determined by 
\begin{equation}
  -4 \eta_{00} + a_0 + 2b_0 - 2 q_{\theta} \equiv 0. 
                                              \qquad ({\rm mod} \ MN) 
\label{eqn:fcm3}
\end{equation}
In order to obtain $\mu \sim \O(10^2)$GeV, 
we need $\eta_{00} \sim 2n$. 
Then, when $b_0$ is even and $a_0 + 2b_0 \leq 0$, 
we rewrite Eq. (\ref{eqn:fcm3}) as 
\begin{equation}
  -4 \eta_{00} + a_0 + 2b_0 - 2 q_{\theta} = -2MN. 
\label{eqn:fc3}
\end{equation}

We now turn to the quark/lepton mass matrices. 
The mass matrix for up-type quarks comes from the term 
\begin{equation}
  m_{ij} \left( \frac {S_0 {\overline S}}{M_S^2} \right)^{\mu_{ij}} 
       \, Q_i U^c_j H_{u0}, \qquad (i, \ j = 1, \, 2, \, 3) 
\end{equation}
with $m_{ij} = \O(1)$. 
The exponent $\mu_{ij}$ is determined by 
\begin{equation}
   -4\mu_{ij} + a_i + b_j + b_0 - 2 q_{\theta} \equiv 0. 
                                              \qquad ({\rm mod} \ MN) 
\label{eqn:fcm4}
\end{equation}
The $3 \times 3$ mass matrix is given by 
\begin{equation}
  {\cal M}_{ij} v_u = m_{ij} \, x^{\mu_{ij}} v_u\,, 
\label{eqn:Mupq}
\end{equation}
where $v_u = \langle H_{u0} \rangle$. 
In order to account for the experimental fact that 
the top-quark mass is of $\O(v_u)$, we expect $\mu_{33} \simeq 0$ 
and set 
\begin{equation}
  -4\mu_{33} + a_3 + b_3 + b_0 - 2 q_{\theta} = 0. 
\label{eqn:fc4}
\end{equation}
This relation holds when $a_3 + b_3 + b_0 \equiv 0$ (mod 4). 
Furthermore, we choose the parameterization 
\begin{equation}
  \a_1 > \a_2 > \a_3 = 0, \qquad \b_1 > \b_2 > \b_3 = 0, 
\end{equation}
where $\a_i$ and $\b_i$ are integers defined by 
\begin{equation}
  \a_i = \frac{1}{4} (a_i - a_3), \qquad 
  \b_i = \frac{1}{4} (b_i - b_3). 
\label{eqn:ab123}
\end{equation}
Then the mass matrix Eq.~(\ref{eqn:Mupq}) is of the form 
\begin{equation}
  {\cal M} \times v_u = \left(
     \begin{array}{ccc}
       m_{11} x^{\a_1 + \b_1}  &  m_{12} x^{\a_1 + \b_2}  &  m_{13} x^{\a_1}  \\
       m_{21} x^{\a_2 + \b_1}  &  m_{22} x^{\a_2 + \b_2}  &  m_{23} x^{\a_2}  \\
       m_{31} x^{\b_1}         &  m_{32} x^{\b_2}         &  m_{33} 
     \end{array}
   \right) \times x^{\mu_{33}} \, v_u. 
\end{equation}
The mass eigenvalues for up-type quarks become 
\begin{equation}
  (m_u, \ m_c, \ m_t) \sim 
  (x^{\a_1 + \b_1}, \ x^{\a_2 + \b_2}, \ 1) \times x^{\mu_{33}} \, v_u 
\end{equation}
at the string scale $M_S$.

In the down-quark sector, the mass matrix is given 
by\cite{Matsu1,Matsu2,Matsu3,CKM} 
\begin{equation}
\begin{array}{r@{}l} 
   \vphantom{\bigg(}   &  \begin{array}{ccc} 
          \quad \,  g^c   &  \quad  D^c  &  
        \end{array}  \\ 
\widehat{{\cal M}}_d  = 
   \begin{array}{l} 
        g   \\  D  \\ 
   \end{array} 
     & 
\left( 
  \begin{array}{cc} 
    y_S {\cal Z}   &     y_N  {\cal M}  \\
      0     &  \rho_d  {\cal M} 
  \end{array} 
\right) 
\end{array} 
\label{eqn:Mhd}
\end{equation}
in $M_S$ units, where $y_S = \langle S_0 \rangle /M_S$, 
$y_N = \langle N^c_0 \rangle /M_S$, $\rho _d = v_d/M_S$ and 
$v_d = \langle H_{d0} \rangle$. 
Since $g^c$ and $D^c$ are indistinguishable under 
the standard model gauge group $G_{\rm SM}$, 
mixings occur between these fields. 
Consequently, the mass matrix for down-type quarks becomes 
a $6 \times 6$ matrix. 
The above $3 \times 3$ $g\,$-$g^c$ submatrix coming from the term 
\begin{equation}
  z_{ij} \left( \frac {S_0 {\overline S}}{M_S^2} \right)^{\zeta_{ij}} 
       \, S_0g_i g^c_j 
\end{equation}
is given by 
\begin{equation}
   {\cal Z}_{ij} = z_{ij} \, x^{\zeta_{ij}}, 
\end{equation}
with $z_{ij} = {\cal O}(1)$. 
Flavor symmetry requires the conditions 
\begin{equation}
  -4\zeta_{ij} + a_i + a_j + a_0 - 2 q_{\theta} \equiv 0. 
                                  \qquad ({\rm mod} \ MN) 
\label{eqn:fcm5}
\end{equation}
Then, with $a_i + a_j \equiv 0$ (mod 4), as shown in Eq.~(\ref{eqn:ab2}), 
we set 
\begin{equation}
  -4\zeta_{33} + 2 a_3 + a_0 - 2 q_{\theta} = 0. 
\label{eqn:fc5}
\end{equation}
The eigenstates of the mass matrix (\ref{eqn:Mhd}) contain 
three heavy modes and three light modes. 
An important phenomenological constraint results from 
the observed pattern of quark mixings. 
As pointed out in Ref. \cite{CKM}, when the relation 
\begin{equation}
    x^{\delta_d} \sim 1
\label{eqn:CKM}
\end{equation}
is satisfied, where 
\begin{equation}
  \delta_d = \left[ \a_1 + \zeta_{33} + \frac{1}{2} \right] - 
    \left[ \b_1 + \mu_{33} + 
    \frac{1}{2} \left( \frac{n}{4} + \delta_N \right) \right], 
\end{equation}
we obtain 
\begin{equation}
  \theta_{12} \sim x^{\a_1-\a_2}, 
        \qquad \theta_{23} \sim x^{\a_2}. 
\end{equation}
By taking $x^{\a_1} \sim \l^3$ and $x^{\a_2} \sim \l^2$ 
with $\l \sim 0.22$, 
we can reproduce a phenomenologically acceptable pattern of 
the CKM matrix : 
\begin{equation}
V_{\rm CKM} \sim \left(
  \begin{array}{ccc}
             1    &   \l    &   \l^5  \\
            \l    &    1    &   \l^2  \\
            \l^3  &   \l^2  &    1 
  \end{array}
  \right). 
\end{equation}
At the string scale $M_S$, the mass spectra of the light modes become 
\begin{equation}
  (m_d, \ m_s, \ m_b) \sim 
  (x^{\a_1 + \b_1 + \delta_d}, \ x^{\a_2 + \b_1}, \ 
       x^{\a_2 + \b_1 - \a_1 + \delta_d}) \times x^{\mu_{33}} \, v_d 
\end{equation}
for $\delta_d \geq 0$, and 
\begin{equation}
  (m_d, \ m_s, \ m_b) \sim 
  (x^{\a_1 + \b_1}, \ x^{\a_2 + \b_1 + \delta_d}, \ 
       x^{\a_2 + \b_1 - \a_1 + \delta_d}) \times x^{\mu_{33}} \, v_d 
\end{equation}
for $\delta_d < 0$.

In the charged lepton sector, the mass matrix has 
the $6 \times 6$ form\cite{Matsu1,Matsu2,Matsu3,MNS} 
\begin{equation}
\begin{array}{r@{}l} 
   \vphantom{\bigg(}   &  \begin{array}{ccc} 
          \quad   H_u^+   &  \quad  E^{c+}  &  
        \end{array}  \\ 
\widehat{{\cal M}}_l = 
   \begin{array}{l} 
        H_d^-  \\  L^-  \\ 
   \end{array} 
     & 
\left( 
  \begin{array}{cc} 
       y_S {\cal H}    &    0       \\
       y_N {\cal M}    &  \rho _d {\cal M} 
  \end{array} 
\right) 
\end{array} 
\label{eqn:Mhcl}
\end{equation}
in $M_S$ units. 
Because $H_d$ and $L$ also have the same quantum number under $G_{\rm SM}$, 
mixings occur between these fields. 
The above $3 \times 3$ $H_d\,$-$H_u$ submatrix coming from 
\begin{equation}
  h_{ij} \left( \frac {S_0 {\overline S}}{M_S^2} \right)^{\eta_{ij}} 
       \, S_0 H_{di} H_{uj} 
\end{equation}
is expressed as 
\begin{equation}
   {\cal H}_{ij} = h_{ij} x^{\eta_{ij}}, 
\end{equation}
with $h_{ij} = {\cal O}(1)$. 
>From the flavor symmetry, we have the conditions 
\begin{equation}
  -4\eta_{ij} + b_i + b_j + a_0 - 2 q_{\theta} \equiv 0. 
                                \qquad ({\rm mod} \ MN) 
\label{eqn:fcm6}
\end{equation}
Again, by assuming $b_i + b_j \equiv 0$ (mod 4), we set 
\begin{equation}
  -4\eta_{33} + 2 b_3 + a_0 - 2 q_{\theta} = 0. 
\label{eqn:fc6}
\end{equation}
The eigenstates of the mass matrix (\ref{eqn:Mhcl}) contain 
three heavy modes and three light modes.

In the neutral sector, there exist five types of matter fields, 
$H_u^0$, $H_d^0$, $L^0$, $N^c$ and $S$. 
Then we have the $15 \times 15$ mass 
matrix\cite{Matsu1,Matsu2,Matsu3,MNS} 
\begin{equation} 
\begin{array}{r@{}l} 
   \vphantom{\bigg(}   &  \begin{array}{cccccc} 
          \quad \, H_u^0   & \quad \  H_d^0  &  \quad \ L^0  
                          &  \quad \ \ N^c   &  \quad \  S  &
        \end{array}  \\ 
\widehat{{\cal M}}_{NS} = 
   \begin{array}{l} 
        H_u^0  \\  H_d^0  \\  L^0  \\  N^c  \\  S  \\
   \end{array} 
     & 
\left( 
  \begin{array}{ccccc} 
       0     &  y_S {\cal H}     &   y_N {\cal M}^T     
                       &      0     &  \rho _d {\cal M}^T  \\
    y_S {\cal H}    &     0      &      0      
                       &      0     &  \rho _u {\cal M}^T  \\
    y_N {\cal M}    &     0     &      0      
                       &  \rho _u {\cal M} &       0       \\
       0     &     0     & \rho _u {\cal M}^T 
                       &      {\cal N}     &      {\cal T}^T      \\
   \rho _d {\cal M} & \rho _u {\cal M} &      0      
                       &      {\cal T}     &       {\cal S}       \\
  \end{array} 
\right) 
\end{array} 
\label{eqn:Mhn}
\end{equation}
in $M_S$ units, where $\rho _u = v_u/M_S$. 
In this matrix, the $6 \times 6$ submatrix 
\begin{equation}
   \widehat{{\cal M}}_M = \left(
   \begin{array}{cc}
        {\cal N}    &   {\cal T}^T  \\
        {\cal T}    &    {\cal S}     
   \end{array}
   \right) 
\end{equation}
plays the role of the Majorana mass matrix in the seesaw mechanism. 
The $3 \times 3$ submatrix ${\cal N}$ is induced from the terms 
\begin{equation}
   M_S^{-1} \left( \frac {S_0 {\overline S}}{M_S^2} \right)^{\nu _{ij}} 
       (\psi_i {\bar \psi})(\psi_j {\bar \psi}), 
            \qquad (i,j = 1,2,3) 
\label{eqn:Majo}
\end{equation}
where the exponents are given by 
\begin{equation}
  -4 \nu_{ij} + b_i + b_j + 2\bar{b} 
      - 2 q_{\theta} \equiv 0. \qquad ({\rm mod} \ MN) 
\label{eqn:fcm7}
\end{equation}
In fact, these terms lead to the Majorana mass terms 
\begin{equation}
   M_S \, {\cal N}_{ij} N^c_i N^c_j \sim M_S \, x^{\nu_{ij}} 
     \left( \frac{\langle \bar{N^c} \rangle}{M_S} \right)^2 N^c_i N^c_j. 
\label{eqn:MajoN}
\end{equation}
Phenomenologically, it is desirable for the Majorana mass of 
the third generation to be $10^{10} - 10^{12}$ GeV. 
This scale is nearly equal to the geometrical average of 
$M_S$ and $M_Z$ : 
\begin{equation}
  M_S x^{\nu_{33}} 
     \left( \frac{\langle \bar{N^c} \rangle}{M_S} \right)^2 
         \sim 10 \times \sqrt{M_S \, M_Z} 
         \sim 50 \times M_S \sqrt{\rho}. 
\end{equation}
This implies 
\begin{equation}
  \nu_{33} + \frac{n}{4} \sim 0.9 \times n. 
\label{eqn:nu33}
\end{equation}
When $b_3$ is even but $\bar{b}$ is odd, 
the flavor symmetry leads to 
\begin{equation}
  -4\nu_{33} + 2 b_3 + 2\bar{b} - 2 q_{\theta} = - MN. 
\label{eqn:fc7}
\end{equation}
Because the right-hand side of this equation is not $-2MN$ but $-MN$, 
we can obtain solutions consistent with Eq. (\ref{eqn:nu33}). 
The submatrix ${\cal S}$ induced from 
\begin{equation}
   M_S^{-1} \left( \frac {S_0 {\overline S}}{M_S^2} \right)^{\sigma _{ij}} 
        (\phi_i {\bar \phi})(\phi_j {\bar \phi}) 
\end{equation}
is expressed as 
\begin{equation}
   {\cal S}_{ij} \sim x^{\sigma_{ij}} 
     \left( \frac{\langle \bar{S} \rangle}{M_S} \right)^2. 
\end{equation}
The exponents are determined by 
\begin{equation}
  -4\sigma_{ij} + a_i + a_j + 2\bar{a} - 2 q_{\theta} \equiv 0. 
                                \qquad ({\rm mod} \ MN) 
\end{equation}
The submatrix ${\cal T}$ induced from 
\begin{equation}
   M_S^{-1} \left( \frac {S_0 {\overline S}}{M_S^2} \right)^{\tau _{ij}} 
(\phi_i {\bar \phi})(\psi_j {\bar \psi}) 
\end{equation}
is given by 
\begin{equation}
   {\cal T}_{ij} \sim x^{\tau_{ij}} \frac{\langle \bar{S} \rangle 
                           \langle \bar{N^c} \rangle}{M_S^2}. 
\end{equation}
The flavor symmetry yields the conditions 
\begin{equation}
  -4\tau_{ij} + a_i + b_j + \bar{a} + \bar{b} 
          - 2 q_{\theta} \equiv 0. \qquad ({\rm mod} \ MN) 
\label{eqn:fcm8}
\end{equation}
Because only $\bar{b}$ is taken as an odd integer, 
we have no solution to satisfy Eq. (\ref{eqn:fcm8}). 
This means that ${\cal T} = 0$.

We now proceed to discuss phenomenological constraints 
resulting from the lepton flavor mixings. 
As pointed out in Ref. \cite{MNS}, in the present framework 
there are two possibilities for realistic patterns of the MNS matrix, 
that is, the LMA-MSW solution and the SMA-MSW solution. 
The LMA solution can be derived when the relation 
\begin{equation}
   x^{\delta_L} \sim 1 
\label{eqn:MNS-L}
\end{equation}
holds, 
where 
\begin{equation}
  \delta_L = \left[ \frac{\a_1 + \a_2}{2} + \mu_{33} 
         + \frac{1}{2} \left( \frac{n}{4} + \delta_N \right) \right] - 
            \left[ \b_1 + \eta_{33} + \frac{1}{2} \right]. 
\end{equation}
In this case we have 
\bea
   \tan \theta_{12}^{\rm MNS} 
          & \sim & x^{\frac{\a_1 - \a_2}{2} + \delta_L} 
            \sim \sqrt{\l} \, x^{\delta_L}, \nonumber \\
   \tan \theta_{23}^{\rm MNS} 
          & \sim & x^{\frac{\a_1 - \a_2}{2} - \delta_L} 
            \sim \sqrt{\l} \, x^{- \delta_L}, \\
   \tan \theta_{13}^{\rm MNS} 
          & \sim & x^{\a_1 - \a_2} \sim \l, \nonumber 
\eea
and the mass spectra of the light charged leptons become 
\begin{equation}
  (m_e, \ m_{\mu}, \ m_{\tau}) \sim 
  (x^{\a_1 + \b_1}, \ x^{ \b_2 + \frac{\a_1 + \a_2}{2} - \delta_L}, \ 
       x^{\a_2}) \times x^{\mu_{33}} \, v_d. 
\end{equation}
The neutrino masses are given by 
\begin{equation}
  (m_{\nu_1}, \ m_{\nu_2}, \ m_{\nu_3}) \sim 
  (x^{2 (\a_1 - \a_2)}, \ x^{ \a_1 - \a_2 - 2 \delta_L}, \ 1 )
       \times \frac{v_u^2}{M_S} 
       x^{2( \a_2 + \mu_{33}) - \nu_{33} - \frac{n}{4} - \delta_N}. 
\end{equation}
In a similar way, the SMA solution is obtained when the relation 
\begin{equation}
   x^{\delta_S} \sim 1 
\label{eqn:MNS-S}
\end{equation}
is satisfied, where 
\begin{equation}
  \delta_S = \left[ \a_1 + \mu_{33} 
           + \frac{1}{2} \left( \frac{n}{4} + \delta_N \right) \right] 
        - \left[ \b_2 + \eta_{33} + \frac{1}{2} \right]. 
\end{equation}
In this case, the parameterizations 
$x^{\b_1} \sim \l^4$ and $x^{\b_2} \sim \l^2$ lead to 
\bea
   \tan \theta_{12}^{\rm MNS} 
          & \sim & x^{\b_1 - \b_2 - \delta_S} 
            \sim \l^2 x^{- \delta_S}, \nonumber \\
   \tan \theta_{23}^{\rm MNS} 
          & \sim & x^{\delta_S}, \\
   \tan \theta_{13}^{\rm MNS} 
          & \sim & x^{\b_1 - \b_2} \sim \l^2. \nonumber 
\eea
We then obtain the mass spectra 
\begin{equation}
  (m_e, \ m_{\mu}, \ m_{\tau}) \sim 
  (x^{\a_1 + 2\b_1 - \b_2 - \delta_S}, \ x^{\a_1 + \b_2}, \ 
       x^{\a_1 - \delta_S}) \times x^{\mu_{33}} \, v_d 
\end{equation}
for light charged leptons and 
\begin{equation}
  (m_{\nu_1}, \ m_{\nu_2}, \ m_{\nu_3}) \sim 
  (x^{2 (\b_1 - \b_2)}, \ x^{ 2 \delta_S}, \ 1 )
       \times \frac{v_u^2}{M_S} 
       x^{2( \a_1 + \mu_{33} - \delta_S) 
               - \nu_{33} - \frac{n}{4} - \delta_N} 
\end{equation}
for neutrinos.

\vspace{10mm}

\section{Anomaly-free conditions}
It is known that all non-gauge symmetries break down 
around the Planck scale due to quantum gravity effects.\cite{Banks} 
On the other hand, phenomenologically it seems that the flavor 
symmetries are necessary for explaining the fermion mass 
hierarchies and the mixings. 
Therefore, it would be natural for the flavor symmetries to be 
unbroken discrete subgroups of local gauge symmteries. 
If this is the case, 
the discrete flavor symmetries would be stable with respect to 
quantum gravity effects and then remains in the low-energy 
effective theory. 
Such discrete flavor symmetries should be 
non-anomalous.\cite{Ib-Ro,KMY}

If the ${\bf Z}_{MN}$ symmetry considered here arises from 
certain gauge symmetries and if anomaly cancellation does not 
occur via the Green-Schwartz mechanism,\cite{G-S} 
the ${\bf Z}_{MN}$ symmetry itself should be non-anomalous. 
Because the gauge symmetry at the string scale is assumed 
to be $SU(6) \times SU(2)_{\rm R}$, 
the mixed anomaly conditions ${\bf Z}_{MN} \cdot (SU(6))^2$ 
and ${\bf Z}_{MN} \cdot (SU(2)_{\rm R})^2$ are imposed on 
the ${\bf Z}_{MN}$ charges of the massless matter fields. 
The heavy fermions decouple in ${\bf Z}_{MN} \cdot (SU(6))^2$ 
and ${\bf Z}_{MN} \cdot (SU(2)_{\rm R})^2$ anomalies 
but not in the cubic ${\bf Z}_{MN}^3$ and 
the mixed ${\bf Z}_{MN} \cdot ({\rm Graviton})^2$ anomalies. 
At present, however, we have no information about the heavy modes. 
Therefore, the cubic ${\bf Z}_{MN}^3$ and 
the mixed ${\bf Z}_{MN} \cdot ({\rm Graviton})^2$ anomaly conditions 
are not relevant to the constraints on the flavor charges of matter fields 
in the low-energy effective theory.

Because the charged matter fields consist of 
$({\bf 15}, \ {\bf 1})$, $({\bf 6^*}, \ {\bf 2})$ and 
their conjugates under $SU(6) \times SU(2)_{\rm R}$, 
the mixed anomaly conditions become 
\bea
  4 \, \left[\sum_{i=0}^3 (a_i - q_{\theta}) 
              + (\bar{a} - q_{\theta}) \right] 
              \phantom{MMMMMMM}   &   &  \nonumber \\
   + \, 2 \, \left[\sum_{i=0}^3 (b_i - q_{\theta}) 
              + (\bar{b} - q_{\theta}) \right]
           + 12 \, q_{\theta} & \equiv & 0, \qquad ({\rm mod} \ MN) \\
  6 \, \left[\sum_{i=0}^3 (b_i - q_{\theta}) 
              + (\bar{b} - q_{\theta}) \right]
           + 4 \, q_{\theta} & \equiv & 0, \qquad ({\rm mod} \ MN) 
\eea
for $SU(6)$ and $SU(2)_{\rm R}$, respectively. 
These conditions are rewritten as 
\begin{equation}
  4 \, a_T + 2 \, b_T \equiv 18 \, q_{\theta}, \qquad 
   6 \, b_T \equiv 26 \, q_{\theta}, \qquad ({\rm mod} \ MN) 
\end{equation}
where 
\begin{equation}
a_T = \sum_{i=0}^{3} a_i + \bar{a}, \qquad 
b_T = \sum_{i=0}^{3} b_i + \bar{b}. 
\end{equation}
Noting that $a_T$ is even and $b_T$ is odd, 
we obtain 
\bea
  a_T - b_T & \equiv & 
        \frac{1}{2}MN - 2 \, q_{\theta}, \qquad ({\rm mod} \ MN) 
\label{eqn:anomc1}        \\
  6 \, a_T     & \equiv & 14 \, q_{\theta}. \ \phantom{MMMMM} ({\rm mod} \ MN) 
\label{eqn:anomc2}
\eea

Because the Grassmann number $\theta$ has charge $(\pm 1, \ 0)$ 
under ${\bf Z}_M \times {\bf Z}_N$, 
in the case $M \equiv 0$ (mod 3), 
we have no solutions of the anomaly condition 
Eq. (\ref{eqn:anomc2}). 
Thus 
\begin{equation}
  M \not\equiv 0. \qquad ({\rm mod} \ 3) 
\end{equation}
In a previous paper\cite{Fuzz}, we chose $M = 15$ and $N = 14$. 
This choice contradicts the above conditions. 
Therefore, in the next section we explore viable solutions 
that are consistent with these anomaly conditions. 
The anomaly conditions are so stringent 
that many types of discrete symmetries are ruled out. 
In fact, as seen in the next section, we find a LMA solution 
but no SMA solution.

Finally, we would like to remark that the 
$D_4={\bf Z}_2^{({\rm F})} \sdpro {\bf Z}_4$ 
mixed anomaly conditions are satisfied in the present model. 
As seen from Tables II and III, under ${\bf Z}_2^{({\rm FC})}$, 
$\phi({\bf 15, \ 1})_i \ (i=0,1,2,3)$ and 
$\bar{\phi({\bf 15, \ 1})}$ are even, 
while $\psi({\bf 6^*, \ 2})_i \ (i=0,1,2,3)$ and 
$\bar{\psi({\bf 6^*, \ 2})}$ are odd. 
Since these fields are even-dimensional representations of 
$SU(6)$ and also of $SU(2)_{\rm R}$, 
the present matter content is anomaly-free with respect to 
the ${\bf Z}_2^{({\rm FC})}$ mixed anomaly. 
For the ${\bf Z}_4$ mixed anomalies, we have to take 
account of the relation 
\begin{equation}
    g_1\, g_2\, g_1^{-1} = g_2^{-1}. 
\end{equation}
Specifically, $g_1$ does not commutate with $g_2$ 
but does commutate with $g_2^2$. 
This relation implies that ${\bf Z}_4$ charges are additive 
not mod 4 but mod 2. 
Therefore, in order to determine whether the ${\bf Z}_4$ 
mixed anomaly conditions are satisfied, 
it is enough to determine whether ${\bf Z}_2^{({\rm R})}$, 
which is a subgroup of ${\bf Z}_4$, is anomalous. 
As shown in Table IV, under ${\bf Z}_2^{({\rm R})}$, 
$\phi({\bf 15, \ 1})_i$ and $\psi({\bf 6^*, \ 2})_i$ $(i=1,2,3)$ 
superfields are odd, 
while $\phi({\bf 15, \ 1})_0$, $\psi({\bf 6^*, \ 2})_0$, 
$\bar{\phi({\bf 15, \ 1})}$ and $\bar{\psi({\bf 6^*, \ 2})}$ 
are even. 
Their fermion components have opposite R-parities. 
Therefore, ${\bf Z}_2^{({\rm R})}$ mixed anomalies of 
$\phi_0(\psi_0)$ and $\bar{\phi}(\bar{\psi})$ cancel 
pairwise with each other.

\vspace{10mm}

\section{Anomaly-free solutions}
In section 3 we studied a set of phenomenological conditions, 
which can be expressed as 
\begin{eqnarray}
   - (b_0 + \bar{b}) = n & = & \frac{1}{4} (MN - q_{\theta}), 
                                            \nonumber  \\
  -4 \zeta_{00} + 3a_0           & = &  2 q_{\theta} ,        
                                            \nonumber  \\
  -4 \eta_{00} + a_0 + 2b_0      & = &  - 8n,                 
                                            \nonumber  \\
  -4 \mu_{33} + a_3 + b_3 + b_0  & = &  2 q_{\theta},         
                                      \label{eqn:pc4}  \\
  -4 \zeta_{33} + 2 a_3 + a_0    & = &  2 q_{\theta},         
                                            \nonumber  \\
  -4 \eta_{33} + 2 b_3 + a_0     & = &  2 q_{\theta},         
                                            \nonumber  \\
  -4 \nu_{33} + 2 b_3 + 2\bar{b} & = &  - 4n + q_{\theta}. 
                                            \nonumber 
\end{eqnarray}
Desirable values of the colored Higgs mass 
and $\mu$ are obtained in the case 
\begin{equation}
   \zeta_{00} \sim 0, \qquad \eta_{00} \sim 2n. 
\end{equation}
The observed fermion mass spectra require parameterizations in which 
$\mu_{33} \sim 0$, $x^{\a_1} \sim \l^3$, $\qquad x^{\b_1} \sim \l^4$ 
and $x^{\a_2} \sim x^{\b_2} \sim \l^2$. 
In order to account for the observed pattern of the CKM matrix, 
we impose the condition 
\begin{equation}
  \zeta_{33} \sim 
     \b_1 - \a_1 + \mu_{33} + \frac{1}{2} \left( \frac{n}{4} - 1 \right). 
\label{eqn:pc8}
\end{equation}
The LMA solution is obtained under the condition 
\begin{equation}
  \eta_{33} \sim \frac{\a_1 + \a_2}{2} - \b_1 
           + \mu_{33} + \frac{1}{2} \left( \frac{n}{4} - 1 \right), 
\label{eqn:MNS-LMA}
\end{equation}
while the condition for the SMA solution becomes 
\begin{equation}
  \eta_{33} \sim \a_1 - \b_2 + \mu_{33} 
                   + \frac{1}{2} \left( \frac{n}{4} - 1 \right). 
\label{eqn:MNS-SMA}
\end{equation}
In addition, from Eq. (\ref{eqn:nu33}) we have the condition 
\begin{equation}
  \nu_{33} \sim \frac{2}{3} n. 
\label{eqn:maj33}
\end{equation}
When $M$, $N$ and $q_{\theta}$ are given, and 
when $\zeta_{00}$, $\eta_{00}$, $\mu_{33}$, $\zeta_{33}$, 
$\eta_{33}$ and $\nu_{33}$ are also given, 
we have too many relations, because there are five undetermined 
${\bf Z}_{MN}$ charges $a_0$, $b_0$, $\bar{b}$, $a_3$ and $b_3$, 
with the seven equations given in Eq.~(\ref{eqn:pc4}). 
The existence of a solution is not certain, 
and proving or disproving its existence is a subtle matter.

As discussed in the previous section, 
the anomaly conditions are given by 
\bea
  a_T - b_T & \equiv & 
        \frac{1}{2}MN - 2 q_{\theta}, \qquad ({\rm mod} \ MN) 
\label{eqn:ano1}        \\
  6 a_T     & \equiv & 14 q_{\theta}. \ \phantom{MMMMM} ({\rm mod} \ MN) 
\label{eqn:ano2}
\eea
>From the parameterization represented by $a_0 + \bar{a} = - 4$, \ 
$b_0 + \bar{b} = - n = - (MN - q_{\theta})/4$ and 
Eq. (\ref{eqn:ab123}), $a_T$ and $b_T$ can be rewritten as 
\begin{equation}
  a_T = 3 a_3 + 4(\a_1 + \a_2) - 4, \qquad 
  b_T = 3 b_3 + 4(\b_1 + \b_2) - n. 
\label{eqn:abT}
\end{equation}
Recalling that $x^{\a_1} \sim \l^3 \sim 10^{-2}$, and so forth, 
and that $x^{2n-1} \sim 2 \times 10^{-17}$, 
we obtain the relations 
\begin{equation}
   \a_1 + \a_2  \sim 0.4 \times n, \qquad 
   \b_1 + \b_2  \sim 0.5 \times n. 
\end{equation}
Solutions of Eqs. (\ref{eqn:ano1}) and (\ref{eqn:ano2}) 
are found only in the case 
\bea
  a_T - b_T & = & \frac{1}{2}MN - 2 q_{\theta}, \qquad  
\label{eqn:ano3}        \\
    a_T     & = & \frac{1}{3} (7 q_{\theta} + 2MN). 
\label{eqn:ano4}
\eea

After some tedious calculations, we find a LMA solution for which 
\bea
   M = 19, \qquad N = q_{\theta} = 18, \qquad n = 81,  \nonumber \\
  a_T = 270, \qquad b_T = 135 \phantom{MMM} 
\eea
and $x^{161} \sim 2 \times 10^{-17}$, $x^{6.3} = \l \simeq 0.22$. 
${\bf Z}_{MN}$ charges ($MN=342$) of the matter fields are listed 
in Table V. 
This parameterization leads to 
\begin{equation}
 (\zeta_{00}, \ \eta_{00}, \ \mu_{33}, \ 
         \zeta_{33}, \ \eta_{33}, \ \nu_{33}) = 
                   (0, \ 158, \ 3, \ 17, \ 2, \ 51). 
\end{equation}
The scales of the colored Higgs mass and $\mu$ are 
\bea
   m_{g_0/g^c_0} & \simeq & \langle S_0 \rangle 
      = x^{0.5} \times M_S \sim M_S,    \\ 
   \mu & \simeq & x^{154.5} \times M_S \sim 100 \ {\rm GeV}. 
\eea
The quark/lepton mass spectra at the scale $M_S$ become 
\bea
   (m_u, \ m_c, \ m_t) & 
           \sim & (\l^{7.8}, \ \l^{5.2}, \ \l^{0.5}) \times v_u, \nonumber \\
   (m_d, \ m_s, \ m_b) & 
           \sim & (\l^{7.8}, \ \l^{6.7}, \ \l^{3.5}) \times v_d, \\
   (m_e, \ m_{\mu}, \ m_{\tau}) & 
           \sim & (\l^{7.8}, \ \l^{5.9}, \ \l^{2.7}) \times v_d \nonumber 
\eea
for $ -\delta_d = \delta_L \sim 1$. 
These results are in accord with a small value of $\tan \b \equiv v_u/v_d$. 
The CKM matrix turns out to be of the form 
\begin{equation}
V_{\rm CKM} \sim \left(
  \begin{array}{ccc}
             1    &   \l    &   \l^5  \\
            \l    &    1    &   \l^2  \\
            \l^3  &   \l^2  &    1 
  \end{array}
  \right), 
\end{equation}
and the mixing angles in the MNS matrix become 
\begin{equation}
 \tan \theta_{12}^{\rm MNS} \sim \l^{0.7}, \qquad 
 \tan \theta_{23}^{\rm MNS} \sim \l^{0.3}, \qquad 
 \tan \theta_{13}^{\rm MNS} \sim \l. 
\end{equation}
The neutrino mass spectra are given by 
\begin{equation}
  (m_{\nu_1}, \ m_{\nu_2}, \ m_{\nu_3}) \sim 
      10^{-1}{\rm eV} \times (\l^{1.9}, \ \l^{0.6}, \ 1 ). 
\end{equation}

\begin{table}[t]
\caption{Assignment of ${\bf Z}_{342}$ charges 
          for matter superfields}
\label{table:V}
\bc
\begin{tabular}{|c|ccc|cc|} \hline \hline 
\vphantom{\LARGE I} & \phantom{M} $\Phi_1$ \phantom{MM} & 
      \phantom{MM} $\Phi_2$ \phantom{MM} & 
        \phantom{MM} $\Phi_3$ \phantom{M} & 
          \phantom{M} $\Phi_0$ \phantom{MM} & 
            \phantom{MM} $\bar{\Phi}$ \phantom{M} \\ \hline
$\phi({\bf 15, \ 1})$   &  $a_1=126$  &  $a_2=102$  &  $a_3=46$  
                                &  $a_0=12$  &  $\bar{a}=-16$  \\
$\psi({\bf 6^*, \ 2})$  &  $b_1=120$  &  $b_2=80$  &  $b_3=16$  
                         &  $b_0=-14$  &  $\bar{b}=-67$  \\ \hline
\end{tabular}
\ec
\end{table}

Unlike the case for the LMA solution, we could not find 
phenomenologically viable SMA solutions in the parameter region 
$MN < 600$ and $m \equiv 0$ (mod 4), 
because it is difficult to realize a situation in which 
the condition (\ref{eqn:maj33}) is compatible with the other conditions. 
Recent experimental data on neutrino oscillations\cite{SNO,Atmos,Solar} 
strongly suggest that the LMA-MSW solution is most favorable.
The result obtained here is consistent with these data.

\vspace{10mm}

\section{Summary and discussion}
In order to construct a string-inspired model that connects 
appropriately with low-energy physics, 
it is of great importance to explore both the gauge symmetry 
and the flavor symmetry at the string scale $M_S$. 
We chose $SU(6) \times SU(2)_{\rm R}$ as the unified gauge symmetry 
at $M_S$. 
The gauge symmetry can be derived from the perturbative heterotic 
superstring theory via the flux breaking. 
The symmetry breaking of $SU(6) \times SU(2)_{\rm R}$ down to 
$G_{\rm SM}$ can take place via the Higgs mechanism without 
matter fields of adjoint or higher representations. 
Because the doublet Higgs and the color-triplet Higgs fields exist in 
different irreducible representations, 
the triplet-doublet splitting problem is solved naturally. 
As the flavor symmetry, we introduced ${\bf Z}_M \times {\bf Z}_N$ 
and the dihedral group $D_4$ symmetries. 
${\bf Z}_M$ and $D_4$ are R symmetries, 
while ${\bf Z}_N$ is a non-R symmetry. 
Introduction of the dihedral group $D_4$ is motivated by 
the phenomenological observation that the R-handed Majorana 
neutrino mass for the third generation is nearly equal to 
the geometrical average of $M_S$ and $M_Z$. 
We assigned the appropriate flavor charges to the matter fields. 
After studying the mixed anomaly conditions, 
we solved them under many phenomenological constraints coming 
from the particle spectra. 
With the stringent anomaly conditions, 
a LMA-MSW solution was found, but no SMA-MSW solution was found. 
The solution includes phenomenologically acceptable results 
concerning fermion masses and mixings and also concerning 
hierarchical energy scales including the GUT scale, 
the $\mu$ scale and the Majorana mass scale of R-handed neutrinos.

We obtained the reasonable particle spectra at an energy scale 
around the scale $M_S$ as shown in the previous section. 
In order to investigate the particle spectra at low-energies, 
we need to study the renormalization-group evolution of 
gauge couplings and the effective Yukawa couplings and to incorporate 
the supersymmetry breaking effect. 
In our LMA-MSW solution, the ratio $m_d/m_e$ at $M_S$ is 
nearly unity, and also we obtain $m_b/m_{\tau} \sim \l$ 
at $M_S$. 
These results are in contrast with those obtained from some conventional 
GUT-type models, in which the ratio $m_b/m_{\tau}$ is predicted 
to be unity at the GUT scale. 
In the present model, we have peculiar particle spectra. 
In particular, there appear colored superfields with 
even R-parity around the TeV region, 
which do not participate in proton decay. 
In the presence of these extra colored particles, 
the $SU(3)_c$ gauge coupling remains almost unchanged in the whole 
region ranging from $M_Z$ to $M_S$. 
Therefore, the renormalization effects of $SU(3)_c$ 
in our model are expected to become rather large compared with those 
in conventional GUT-type models. 
Thus it seems that the particle spectra at $M_S$ obtained here 
are consistent with those at low energies. 
A detailed study of the renormalization group evolution will 
be presented elsewhere.

In this paper we assumed that the flavor symmetry contains 
the semi-direct product group $D_4$, which is an extension 
of R-parity. 
It would be interesting to explore other possibilities for 
the semi-direct product flavor symmetry. 
Among them we may find more simple flavor symmetries, which 
could lead to phenomenologically viable results.

\section*{Acknowledgements}
Two of the authors (M. M. and T. M.) are supported in part by 
a Grant-in-Aid for Scientific Research, 
Ministry of Education, Culture, Sports, Science and Technology, 
Japan (No.~12047226).



\end{document}